# Infinite-layer nickelate superconductors:

# A current experimental perspective of the crystal and electronic structures

L. E. Chow and A. Ariando[*]

Department of Physics, Faculty of Science, National University of Singapore, Singapore 117551, Singapore

*To whom correspondence should be addressed: ariando@nus.edu.sg

## Abstract

After the reward of more than two decades of pursuit on the high-$T_c$ cuprate analog with the hope to obtain a better understanding of the mechanism of high-$T_c$ superconductivity, the discovery of superconductivity in the infinite-layer nickelate brings more mystery to the picture than expected. Tops in the list of questions are perhaps (1) absence of superconductivity in the bulk nickelate and limited thickness of the infinite-layer phase in thin film, (2) absence of superconductivity in the La-nickelate despite it being the earliest studied rare-earth nickelate, and the role of $4f$ orbital in the recipe of superconductivity, (3) absence of Meissner effect and suspect of the origin of superconductivity from the interface, (4) whether nickelate hosts similar pairing symmetry to the single-band high-$T_c$ cuprates or multiband iron-based superconductor.

In this perspective article, we will discuss the following aspects: (1) stabilization of the infinite-layer phase on the $SrTiO_3$(001) substrate and the thickness dependency of observables; (2) rare earth dependence of the superconducting dome and phase diagram on the (La/Pr/Nd)- infinite-layer nickelate thin film; (3) experimental aspects of the measurement of Meissner effect; (4) theoretical framework and experimental study of the pairing symmetry of infinite-layer nickelate superconductor.

**Main text**

Around four decades ago, the witness of superconductivity above 30K redefined preexisting knowledge on the mechanism of superconductivity and restructured the landscape of the playground on superconductor materials [1,2]. Understanding the high-temperature (high-$T_c$) superconductivity has since been one of the holy grails in physics. Several characteristic properties were discussed in the cuprate superconductor: (1) quasi-2D $CuO_2$ square-planar lattice, (2) antiferromagnetic order and superexchange interaction, (3) spin $S = \frac{1}{2}$ half-filling state, (4) $Cu^{2+}$ of $3d^9$ electronic configurations [3–7]. To identify which parameters are the key ingredients which drive the high-$T_c$ superconductivity in cuprate, searching for isostructural compounds with some of these properties and comparing them to the cuprate were motivated [8]. Among all the cuprate analogs [9], nickelate of $Ni^{1+}$ with the same $3d^9$ electronic configurations was identified as the closest cousin to the cuprate [7,10]. For decades, theoretical and experimental efforts have been made to explore the lead [11–13]. Unlike cuprate, which was first synthesized in the bulk form, superconducting nickelate was only realized recently in the thin-film form [14–21], with $Ni^{1+}$ in the infinite-layer phase that can be achieved through topotactic reduction from the perovskite compound. Missing superconductivity in the bulk nickelate [22,23] and limitation in stabilizing the infinite-layer phase above ~10 nm from the substrate [24–26] demands answers on the thickness-dependent crystallinity and electronic structure of the infinite-layer nickelate film. While lanthanide-cuprate La-Ba-Cu-O was the first superconducting compound synthesized [27], lanthanide-nickelate was initially reported to not host superconductivity, but superconductivity was only realized in the neodymium-based counterpart, a rare-earth element with $4f$ magnetism [14]. The possible roles of rare-earth magnetism in the early observation of nickelate's superconductivity added layers of mystery to the newfound sister of cuprate [6,28,29]. In addition to the bulk nickelate not being reported to show superconductivity, concrete evidence of the Meissner effect in the superconducting nickelate thin film was missing [14], leading to the suspect whether the phenomenon was interfacial in nature. High-$T_c$ cuprate is uniquely identified with a dominant $d_{x^2-y^2}$-wave gap which is believed to be mediated by the antiferromagnetic superexchange interaction, that poses as a crucial factor in the high-$T_c$ superconductivity [3]. Hence, answering the superconducting order parameter in the nickelate is the top priority. However, the challenge in the fabrication of high-quality infinite-layer nickelate films obstructed the experimental means to investigate, especially with those surface-sensitive techniques are not applicable when bad crystallinity or secondary phases are prone to form at the surface of nickelate superconducting thin-film [25]. Overall, while the discovery of superconductivity in the nickelate provided an exciting playground to study the highly correlated system, the newfound superconductor family also ignited controversial debates that will reshape high-$T_c$ framework [4–6,30–36]. In this article, we provided a contemporary experimental perspective on the topics.

## 1. Thickness dependence

*1.1. Stabilization of infinite-layer phase*

Since the observation of superconductivity in the infinite-layer nickelate $Nd_{0.8}Sr_{0.2}NiO_2$ thin film [14], apparent challenges have emerged in material synthesis [25]. On top of low reproducibility and difficulty in the fabrication of superconducting doped infinite-layer structure by many experimental groups [37,38], a hard-nut-to-crack issue is a limited thickness from the substrate interface, which the infinite-layer phase can be stabilized [25,39,40]. Above ~10 nm from the substrate interface, obvious secondary phases or off-tilt structures form instead of the infinite-layer or partially reduced perovskite phase [25]. In many cases, growing a thick film will lead to the formation of a secondary phase even at just ~5 nm from the substrate (Figure 1a) [40]. On the other hand, if a thinner <10 nm film is grown, the entire film can be fully reduced with no observation of secondary phase even at the film surface (Figure 1c,d), except for possibly Ruddlesden-Popper (RP) stacking fault at some regions of the film (Figure 1d). The obvious strategy for obtaining the purest possible infinite-layer phase is fabricating thin films below 10 nm. Some reports suggested using $SrTiO_3$ (STO) capping layer on top of the nickelate thin film prior to a topotactic reduction that can serve as a 'backbone' to help stabilize the infinite-layer phase during topotactic reduction from the perovskite phase [25,41]. However, such a method has not led to a thicker infinite-layer phase of >10 nm. Lattice coherency of a crystalline thin film can be seen from the X-ray diffraction (XRD) Laue fringes which originate from the constructive interference between perfect lattice layers of the thin film. To date, most reported XRD data of the perovskite phase of the doped nickelate thin film has clear Laue fringes in the vicinity of the (002) peak; however, Laue fringes are typically absent for the reduced doped infinite-layer nickelate (002) peak [16,40–43]. This may suggest the presence of nonstoichiometric oxygen at random parts of the reduced infinite-layer thin film. Significant development of secondary or perovskite phases is typically avoided after optimization in film growth conditions since no perovskite peak or defect phase peak is seen in the XRD curve.

The challenge of obtaining a coherent infinite-layer phase does not affect transport-related study since zero resistance can be observed even when only a small part of the film is superconducting. Unfortunately, the same cannot be said for measurements requiring a pure phase with coherent crystallinity, especially at the film surface, such as the Angle-resolved Photoemission Spectroscopy (ARPES) or Scanning Tunneling Spectroscopy (STS). Many probing techniques which reveal crucial aspects of the superconductivity in nickelate cannot be carried out because of the lack of coherent lattice and purity in the infinite-layer phase, especially on the top surface. With much effort in optimizing film quality, we recently reported an observation of clear Laue fringes in the vicinity of XRD (002) peak of the infinite-layer phase for $Nd_{1-x}Sr_xNiO_2$ and $La_{1-x}Ca_xNiO_2$ (Figure 1b). Especially in the case of superconducting lanthanide infinite-layer nickelate, more than 30 unit-cells (uc) of a coherent infinite-layer lattice can be seen vividly from the XRD Laue fringes.

*1.2. Thickness dependency and role of strain and interface*

Superconductivity is missing in the bulk infinite-layer nickelate [23,44–46]. The puzzling limitation in the thickness of the infinite-layer thin film further warrants the importance of investigating the thickness dependency of the physical observables and electronic structure of the infinite-layer nickelate. The zero-resistivity critical temperature, $T_{c,0}$, of $Pr_{0.8}Sr_{0.2}NiO_2$ thin film with thickness from 5.3 nm to 12 nm has been reported [8], showing a slight decrease from 5.3 nm to 8 nm and it then increases again up to 12 nm. However, the normal state resistivity of the same samples also shows corresponding change, where high $T_{c,0}$ samples have low resistivity. This implies the stronger correlation between $T_{c,0}$ and normal state resistivity but not the film thickness [25]. On the other hand, the $T_{c,0}$ of $Nd_{0.8}Sr_{0.2}NiO_2$ thin films monotonically increases with film thickness (4.6 nm – 10.1 nm) observed from both resistivity and susceptibility measurements [24]. In addition, a systematic evaluation of the thickness-dependent electronic structure has been shown with the change in Hall coefficients and XAS spectra of $Nd_{0.8}Sr_{0.2}NiO_2$ thin film from 4.6 nm to 10.1 nm [24]. The results imply strain modulation and interface effect in the infinite-layer nickelate.

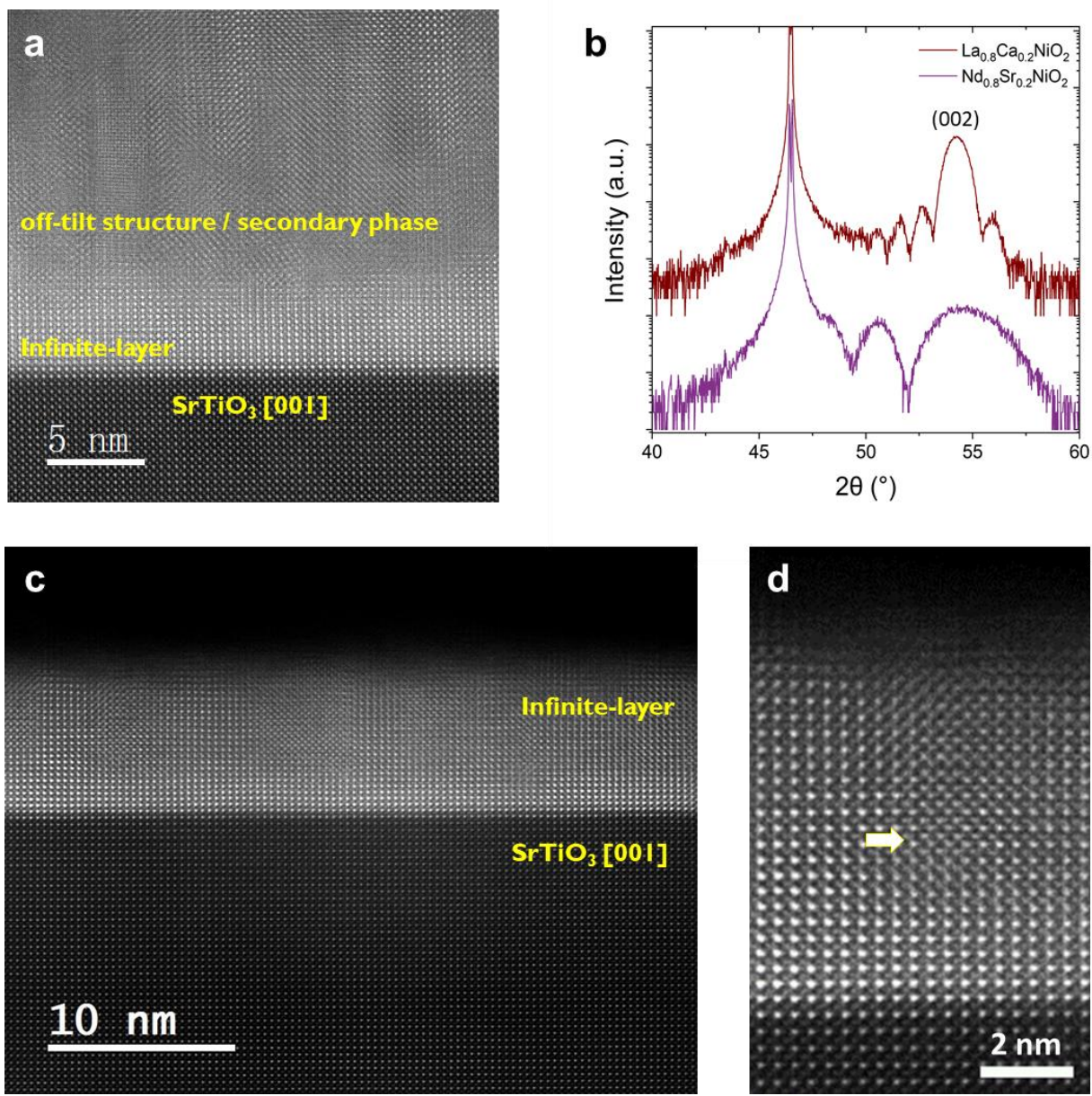

**Figure 1:** Structural properties of the reduced infinite-layer nickelate thin film. (**a, c, d**) STEM HAADF images. (**b**) XRD (002) pattern. (**a**) A thick 30 nm $Nd_{0.85}Sr_{0.15}NiO_2$ film has infinite-layer phase stabilized up to ~5 nm from the substrate interface. Above 5 nm, secondary phase or off-tilt structure can be seen, data is adapted from [40]. Entire $Nd_{0.8}Sr_{0.2}NiO_2$ film of thinner <10 nm can be reduced to infinite-layer phase with perfect crystallinity (**c, d**) up to the surface, with no sign of secondary phase except for possible RP stacking fault occurs at some part of the film (**d**). (**b**) A clear Laue fringes can be observed for 8 nm $Nd_{0.8}Sr_{0.2}NiO_2$ and 15 nm $La_{0.8}Ca_{0.2}NiO_2$ thin films, indicating coherent infinite-layer phase.

## 2. Rare-earth dependence

*2.1. Doping dependent phase diagram*

The barium (Ba) hole-doped lanthanide La-cuprate $La_{2-x}Ba_xCuO_4$ was the first high-$T_c$ cuprate synthesized, which kicked off the door to high-temperature superconductivity beyond the BCS paradigm [27]. To mimic cuprate's square-planar structure and $3d^9$ electronic configurations, $Ni^{1+}$ state in nickelate was predicted to be an ideal cuprate analog to assist in the understanding of the origin of high-$T_c$ superconductivity in cuprate [4,7,13,14,34,47,48]. Naturally, lanthanum (La) was the first rare-earth option to be looked for in nickelate. Many experimental attempts were made on La-nickelate to achieve superconductivity with $Ni^{1+}$ state in the form of superlattices, infinite-layer structure for the past two decades [12,14,49,50]. Despite the long search, La-nickelate was initially found not to be superconducting and the first observation of nickelate superconductivity was realized by Sr-doped on a smaller neodymium ion Nd-infinite-layer nickelate thin film with $4f$ magnetism in 2019 [14]. Since then, superconductivity in the infinite-layer nickelate family has been quickly expanded to another neighbor with rare-earth magnetism in the rare-earth series, praseodymium (Pr) [42], and has been successfully reproduced by multiple experimental groups [17,51]. However, superconductivity in La-nickelate, which has an empty $4f$ orbital, still seems non-existent for another two years until it was successfully realized independently by two different groups in the Ca-doped $La_{1-x}Ca_xNiO_2$ [16] and Sr-doped $La_{1-x}Sr_xNiO_2$ [41] infinite-layer thin film. Given the missing report of superconductivity in La-nickelate for almost two decades, the role of $4f$ magnetism and other differences between $LaNiO_2$ and $(Pr/Nd)NiO_2$ in the recipe of superconductivity become an open question and inspire further investigation on their pairing symmetries, anisotropy and doping dependent phase diagram [16,28,29,39–41,52–55].

Figure 2a shows doping-dependent superconducting dome and phase diagram in various Sr-doped and Ca-doped rare-earth (La, Pr, Nd) infinite-layer nickelate synthesized so far [15,16,39–41]. The first eye-catching feature is the presence of 'dip' with a lower $T_c$ for a particular doping in the $Nd_{1-x}Sr_xNiO_2$ [39,40] and $La_{1-x}Sr_xNiO_2$ [41] thin films. Given the lack of consistency of the 'dip' feature on a particular doping across different rare-earth nickelates and reports, it is presently unclear whether it is an experimental artifact or a reminiscence of certain quantum critical transitions in the system. The second observation is the expansion of the superconducting dome to a higher doping level from $Nd_{1-x}Sr_xNiO_2$, $Pr_{1-x}Sr_xNiO_2$ to $La_{1-x}Ca_xNiO_2$. Given the increased difficulty in synthesizing infinite-layer nickelate thin film at larger doping [25,41], such observation may have a certain correlation to film crystallinity. The perovskite nickelate has a smaller in-plane lattice constant than the $SrTiO_3$ (001) substrate, which the lattice mismatches are smaller from $NdNiO_3$ to $LaNiO_3$. It has been established that a perovskite phase nickelate with good crystallinity is essential to the success in topotactic reduction to the infinite-layer phase [25]. One may suggest that perovskite $La_{1-x}Ca_xNiO_3$ and $Pr_{1-x}Sr_xNiO_3$ can be grown to have better crystallinity than the $Nd_{1-x}Sr_xNiO_3$ at a large doping regime. A similar argument could be made for the difference in superconducting dome between $La_{1-x}Ca_xNiO_2$ and $La_{1-x}Sr_xNiO_2$. The $Ca^{2+}$ ion is of more similar size to the rare earth $La^{3+}$, $Pr^{3+}$ and $Nd^{3+}$ ions than the larger $Sr^{2+}$ ion [16]. Regardless, it is also possible that different rare-earth ion and cation doping leads to a slight difference in the electronic band structure of the infinite-layer nickelate, causing a different span of superconducting domes. Another note is, so far, the doping level in the fabricated thin films is expected to follow the stoichiometry of the polycrystalline target used in the pulsed-laser-deposition (PLD) growth. It is not warranted that the doping level is accurate, and further investigation on the stoichiometry of the doped thin film shall be carried out. The third observation is the correlation between the size of the rare-earth ions and superconducting transition temperature $T_c$. The maximum onset temperature in resistivity, where resistivity reaches 90% of its value at 20K, $T_{c,90\%}$ is roughly consistent among various reports, where La-nickelate has $T_{c,90\%}$ ~9 – 10K while Pr- and Nd-nickelate has $T_{c,90\%}$ ~12 – 15K. It is routinely explained by the increase in electronic bandwidth for a smaller rare-earth ion [14,16].

In addition to the difference in the superconducting $T_c$ between La- and (Pr/Nd)- infinite-layer nickelate, the Hall coefficient ($R_H$) sign change temperature across doping levels also exhibits dissimilarity between La- and (Pr/Nd)-nickelate. $R_H$ sign change temperature is monotonically increasing with hole doping in the case of (Pr/Nd)- infinite-layer nickelate [15,39,40], which may suggest increased-dominancy in $d_{x^2-y^2}$ hole pocket at Fermi level with increasing hole doping. However, in the case of $La_{1-x}Ca_xNiO_2$, the $R_H$ sign change temperatures are constant at around 35K between $0.23 \leq x \leq 0.3$ and the difference in $R_H$ at low temperature is small at increasing doping [16]. While preliminary and possibly confounded by the impact of in-plane compressive stress from the substrate, the role of hole doping is likely to be different in modifying the band structure of the Ca-doped La-nickelate as compared to the Sr-doped (Pr/Nd)-nickelate.

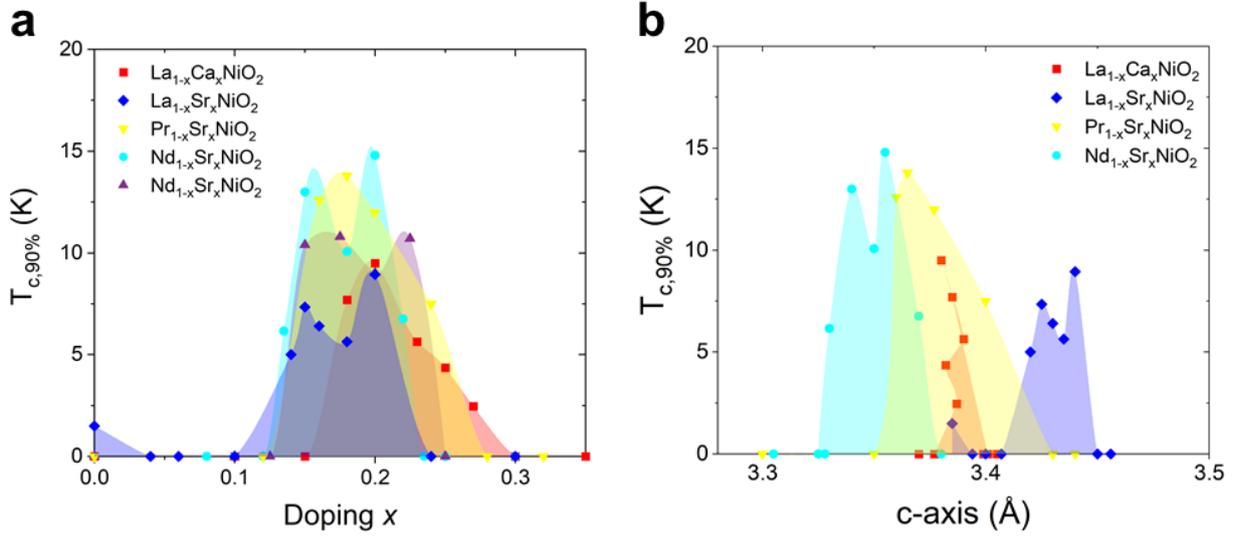

**Figure 2:** Superconducting phase diagram of various doped infinite-layer nickelates, plotted as a function of hole doping levels $x$ (**a**) and c-axis lattice constant (**b**). Data are adapted from (La, Ca) [16], (La,Sr) [41], (Pr, Sr) [15], (Nd, Sr) [39,40].

*2.2. Relevance to lattice constant*

Since superconductivity has not been observed in the bulk infinite-layer nickelate despite that high crystallinity samples were made [44,45], the ab-axis lattice constants of the superconducting infinite-layer thin film are epitaxially constrained to the substrate in-plane lattice constants. It might be intuitive to look for any correlation between the c-axis lattice dimension and superconducting dome of the infinite-layer nickelate (Figure 2b). At first glance, the superconducting dome is generally limited to between 3.32 Å and 3.45 Å. However, the superconducting domes between various rare-earth compounds do not completely overlap. Also, for the case of $La_{1-x}Ca_xNiO_2$, the variation in c-axis lattice constant across doping is small due to a very similar ionic size between $Ca^{2+}$ ion and $La^{3+}$, and has slight sample-to-sample variation at the same Ca doping [16]. It seems early to suggest any strict correlation between the superconducting dome of infinite-layer nickelate and the c-axis lattice dimension. It is worth noting that an enhancement in onset $T_c$ was realized experimentally by applying external pressure [17], which suggests the likelihood of further increasing T$_c$ by simulating chemical pressure through tuning in-plane lattice constants, rare-earth ions size, or dopant size. In addition, we note that there are recent theoretical calculations on the effect of in-plane lattice constant and epitaxial strain in tuning the *P4/mmm - I4/mcm* phase transition in RNiO$_2$ (R=La, Pr, Nd, Eu-Lu, Y) [56,57]

## 3. Observation of Meissner effect

Two main phenomena characterize superconductivity: (1) zero electrical resistivity, (2) Meissner effect. The first experimental report of the discovery of superconductivity in Nd$_{0.8}$Sr$_{0.2}$NiO$_2$ thin film provided multiple resistivity-temperature ($R-T$) curves with clear zero resistivity data and two-coil mutual inductance measurement to observe the expulsion of the magnetic field in the Meissner state [14]. However, the real part of the pickup voltage $Re(V_p)$ does not go to zero as expected for a superconductor in the Meissner state. Some diamagnetic signal is observed, but the $Re(V_p)$ measured the lowest temperature is far from zero as compared to the change within the transition $Re(V_p) \sim 1.7 \rightarrow 1.25~\mu V$ [14]. The absence of concrete data to support the presence of Meissner state led to a suspicion that the superconductivity in nickelate arises from the interface with the substrate but is not intrinsic to the bulk material.

Meissner effect is routinely seen in the magnetic susceptibility measurement: (1) a negative slope in the $M-H$ curve which ends at the lower critical field $H_{c1}$, (2) negative diamagnetic signal below T$_c$ in $M-T$ curve which the volume susceptibility $\chi_V$ (in S.I. unit) goes to -1 for Meissner state. The $M-T$ and $M-H$ data were not presented in the early reports of the superconductivity observed in the infinite-layer nickelate thin films. Zeng *et al.* provided a study on the thickness-dependent effect on the film's T$_c$ in both resistivity and susceptibility measurements [24]. The diamagnetic moment in $M-T$ curve (measured at $H \parallel c$) below the superconducting transition is typically of $\sim 10^{-5}$ emu scale for a $2.5 \times 5$ mm$^2$ superconducting thin film of 8 nm thick. After demagnetizing field correction, the volume susceptibility at 2K can be calculated to be $\chi_V < -0.9$ which is fairly close to -1 for perfect diamagnetism (Figure 3a). Figure 3b presents the raw data for the magnetic moment measured above transition $T > T_{c,onset}$. The infinite-layer thin film on STO substrate has an almost temperature-independent moment for $T_{c,onset} < T < 300K$, which the small ~$10^{-8}$ emu negative moment measured shall correspond to the STO substrate diamagnetic signal. In addition, a negative slope in $M - H$ curve was also observed in superconducting Nd$_{0.8}$Sr$_{0.2}$NiO$_2$ thin film, resembling the Meissner effect and bulk superconductivity, which is intrinsic to the nickelate thin film but not of the interface. The lower critical field $H_{c1}$ is defined as the magnetic field in the sample which the field penetrates the sample volume and the onset of mixed state for type II superconductor. In the $M - H$ curve, due to the thin-film nature with large demagnetizing factor $N \to 1$ for $H \parallel c$, the $H_{c1}^{applied}$ before demagnetization factor correction is around and less than 1 Oe, which is difficult to be resolved and has a large error in the calculation of actual $H_{c1}$. After demagnetizing factor correction, the $H_{c1}(T = 0K)$ is approximated to be around 79 Oe.

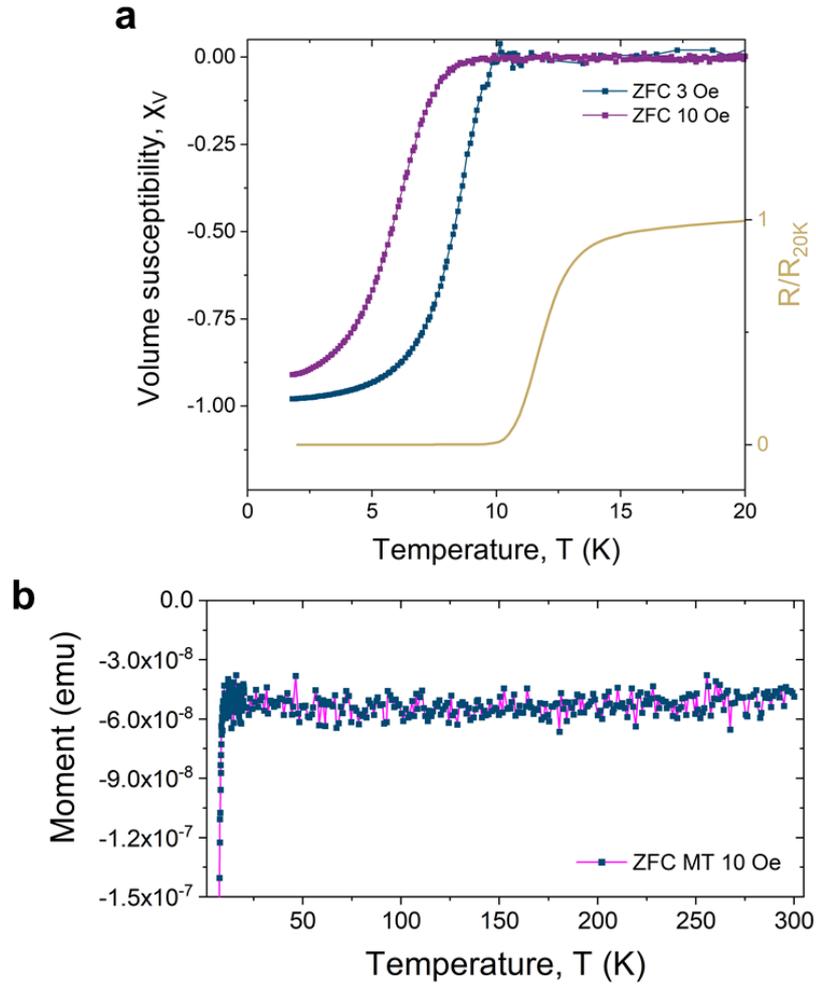

**Figure 3:** Observation of Meissner effect in the superconducting infinite-layer thin film at $H \parallel c$. Superconducting transition observed in resistivity and susceptibility (**a**) measurements for Nd$_{0.8}$Sr$_{0.2}$NiO$_2$ thin film. (**b**) The measured moment at above $T_c$ is typically of $-10^{-8}$ emu scale and almost constant at $T_{c,onset} < T < 300K$, correspond to STO substrate diamagnetic signal.

The superconducting transition temperature observed in the $M - T$ curve typically has an onset close to the $T_{c,0}$ in resistivity provided good homogeneity of the entire sample (Figure 3a). Lower $T_{c,onset}$ in $M - T$ can be measured if the applied field is larger than the lower critical field $H > H_{c1}$. Since the applied $H_{c1}^{applied} \leq 1$ Oe for $Nd_{0.8}Sr_{0.2}NiO_2$ infinite-layer nickelate thin film, a larger measuring field of 10 Oe can lead to lower $T_{c,onset}$ and smaller $\chi_V(2K)$ as compared to 3 Oe measuring field, for example. In addition, the Meissner effect will not be observable, or only a very small diamagnetic signal is observed if there are inhomogeneity and defect phases in the infinite-layer nickelate thin film, which can easily present even when a high $T_c$ is observed in the resistivity $R - T$ data.

## 4. Pairing symmetry

### 4.1. Dominant d-wave gap

The infinite-layer nickelate is a sister of the high-T$_c$ cuprate, which hosts a dominant $d_{x^2-y^2}$-wave gap, both sharing a similar crystal structure and $3d^9$ electronic configuration. In the recent resonant inelastic X-ray scattering (RIXS) experiments at Ni $L_3$-edge on the infinite-layer nickelate thin films, charge order and spin wave of antiferromagnetically coupled spins in a square lattice was observed [43,58–60]. The antiferromagnetic exchange coupling strength $J$ is estimated to be around 63.6 meV in RIXS. While a different $J$ value can be estimated with different spin model calculation, the general consensus is that the nickelate's $J$ value is smaller than the $J\sim130$ meV of the high-T$_c$ cuprates [58,61]. In addition, exchange bias effect was observed at a ferromagnet/Nd$_{0.8}$Sr$_{0.2}$NiO$_2$(20 nm) interface which could be interpreted as the antiferromagnetic nature at the surface of thick Nd$_{0.8}$Sr$_{0.2}$NiO$_2$ film (though exchange bias field is absent for ≤10 nm film) [18]. Despite the lack of concrete proof on the long-range magnetic order to date, the $t - J$ model used in cuprates is perceived to be suitable to capture nickelate's superconductivity [55] and the general consensus across different theoretical calculations on the superconducting pairing symmetry of nickelate is a dominant $d_{x^2-y^2}$-wave pairing like the cuprate with some pointed out various possibility of multiband superconductivity [35,52,55,62]. In the $t - J - K$ model which accounted for the Kondo coupling in nickelate, an interstitial $s$-wave gap exists at the large hole doping and small $t/K$ region of the phase diagram [63]. On the other hand, if hopping $t/K$ is large as compared to the Kondo coupling, a dominant $d$-wave pairing or a transition from $(d + is)$-wave at low doping to $d$-wave at large doping is expected [63].

Considering the role of nickelate superconductivity in illuminating the origin of high-$T_c$ superconductivity in cuprates, a detailed experimental study on the pairing symmetry of nickelate is crucial. The first experimental report on the superconducting gap symmetry of the infinite-layer nickelate is a single-particle-tunneling experiment on $Nd_{0.8}Sr_{0.2}NiO_2$ film surface which detected signals correspond to $s$-wave, $d$-wave or a mixture of both in different parts of the film surface [19]. However, we note here the difficulty in achieving a good crystallinity and high purity of the superconducting phase, especially near the film surface of the reduced infinite-layer thin film. Hence, surface-sensitive techniques which are useful in determining the gap profile like the angle-resolved photoemission spectroscopy (ARPES) may not be feasible to investigate the pairing order of the infinite-layer nickelate thin films until the film quality at the surface is perfected. Furthermore, phase-sensitive experiments are also waiting to be seen [64].

*4.2. Fully gapped pairing and isotropic upper critical field*

While the tunneling experiment did not lead to a complete answer of the nickelate's pairing order, the existence of a fully gapped $s$-wave signal ignited multiple explanations to the observation [63,65]. Recently the upper critical field $H_{c2}$ of $Nd_{0.775}Sr_{0.225}NiO_2$ thin film is measured to be mostly isotropic down to the lowest temperature and is smaller than the Pauli limit [66]. The isotropic $H_{c2}$ in the $Nd_{0.775}Sr_{0.225}NiO_2$ infinite-layer nickelate is completely distinct from the cuprate with large anisotropy and other quasi-2D superconductors. On the other hand, this isotropic upper critical field behavior places nickelate to be more similar to the high-$T_c$ iron-based superconductor which is believed to host nodeless $s^{\pm}$-wave multigap pairing [67,68]. Further investigation on the nickelate's controversial pairing symmetry is warranted.

# Acknowledgement

We thank S.W. Zeng for the fruitful discussion and revision of this article. We acknowledge Elbert E.M. Chia, C.Z. Diao, W. Escoffier, M. Goiran, S.K. Goh, J.X. Hu, H. Jani, Z.S. Lim, C.J. Li, P. Nandi, G.J. Omar, M. Pierre, D. Preziosi, S.K. Sudheesh, M. Salluzzo, C.S. Tang, Andrew T. S. Wee, K.Y. Yip, X.M. Yin, P. Yang, Z.T. Zhang for discussions. This research is supported by the Agency for Science, Technology, and Research (A*STAR) under its Advanced Manufacturing and Engineering (AME) Individual Research Grant (IRG) (A1983c0034) and the Singapore National Research Foundation (NRF) under the Competitive Research Programs (CRP Grant No. NRF-CRP15-2015-01).